\input harvmac
\input tables
\def\O{{\cal O}}
\def\ra{\rightarrow}
\def\dKM{\delta_{\rm KM}}
\def\dkm{\delta_{\rm KM}}
\def\lam{\lambda}
\def\apks{a_{\psi K_S}}
\def\apnn{a_{\pi\nu\bar\nu}}
\def\epsK{\varepsilon_K}
\def\epsH{\epsilon_H}
\def\epsCP{\epsilon_{CP}}
\def\epsa{\epsilon_1}
\def\epsb{\epsilon_2}
\def\epsc{\epsilon_3}
\def\epsd{\epsilon_4}
\def\gsim{{~\raise.15em\hbox{$>$}\kern-.85em
          \lower.35em\hbox{$\sim$}~}}
\def\lsim{{~\raise.15em\hbox{$<$}\kern-.85em
          \lower.35em\hbox{$\sim$}~}}

\def\YGTitle#1#2{\nopagenumbers\abstractfont\hsize=\hstitle\rightline{#1}%
\vskip .4in\centerline{\titlefont #2}\abstractfont\vskip .3in\pageno=0}
\YGTitle{WIS-98/2/Jan-PP, hep-ph/9801411}
{\vbox{\centerline{Approximate CP in Supersymmetric Models}}}
\bigskip
\centerline{Galit Eyal$^{a,b}$ and Yosef Nir$^a$}
\smallskip
\centerline{\it $^a$Department of Particle Physics,
 Weizmann Institute of Science, Rehovot 76100, Israel}
\smallskip
\centerline{\it $^b$Department of Physics, Technion -- Israel
Institute of Technology, Haifa 32000, Israel}
\bigskip
\bigskip
\baselineskip 18pt
\noindent
We construct phenomenologically viable supersymmetric models where CP is
an approximate symmetry. The full high energy theory has exact CP
and horizontal symmetries that are spontaneously broken with a naturally
induced hierarchy of scales, $\Lambda_{CP}\ll\Lambda_H$. Consequently,
the effective low energy theory, that is the supersymmetric Standard
Model, has CP broken explicitly but by a small parameter. The $\epsK$
parameter is accounted for by supersymmetric contributions. The
predictions for other CP violating observables are very different from
the Standard Model. In particular, CP violating effects in neutral $B$
decays into final CP eigenstates such as $B\ra\psi K_S$ and in
$K\ra\pi\nu\bar\nu$ decays are very small.
\bigskip

\baselineskip 18pt
\leftskip=0cm\rightskip=0cm

\Date{}

\newsec{Introduction}

Within the Standard Model, the following features regarding
CP violation hold:
\item{(i)} CP is broken explicitly.
\item{(ii)} All CP violation arises from a single phase
(that is the Kobayashi-Maskawa phase $\dKM$).
\item{(iii)} The measured value of $\epsK$ requires that $\dKM$
is of order one. (In other words, CP is not an approximate symmetry of
the Standard Model.)
\item{(iv)} The values of all other CP violating observables
can be predicted. In particular, the asymmetry $\apks$,
\eqn\defapks{\apks\sin(\Delta m_B t)=-
{\Gamma(B^0_{\rm phys}(t)\ra\psi K_S)
-\Gamma(\bar B^0_{\rm phys}(t)\ra\psi K_S)\over
\Gamma(B^0_{\rm phys}(t)\ra\psi K_S)
+\Gamma(\bar B^0_{\rm phys}(t)\ra\psi K_S)},}
(and similarly various other CP asymmetries in
$B$ decays), and the ratio $\apnn$,
\eqn\defapnn{\apnn={\Gamma(K_L\ra\pi^0\nu\bar\nu)\over
\Gamma(K^+\ra\pi^+\nu\bar\nu)},}
are expected to be of order one.

The commonly repeated statement that CP violation is one of the least
tested aspects of the Standard Model is well demonstrated by the fact
that none of the above features necessarily holds in the presence
of New Physics. Such a dramatic difference from the Standard Model
is possible, for example, in the supersymmetric framework. (For a recent
review of CP violation in supersymmetry, see
\ref\GNR{Y. Grossman, Y. Nir and R. Rattazzi, hep-ph/9701231, to appear
in {\it Heavy Flavours II}, eds. A.J. Buras and M. Lindner (World
Scientific).}.)
Indeed, in this work, we construct phenomenologically viable
supersymmetric models, with the following features:
\item{(i)} CP is an exact symmetry of the full theory but is
spontaneously broken at some high energy scale
by a VEV of a gauge singlet scalar field.
\item{(ii)} In the low energy effective theory, there are many
independent CP violating phases, in particular in the mixing matrices
of gaugino couplings to fermions and sfermions.
\item{(iii)} In the low energy effective theory, CP is an approximate
symmetry. The Kobayashi-Maskawa phase is too small to account for
$\epsK$ which is explained, instead, by supersymmetric contributions.
\item{(iv)} The values of all other CP violating observables can
be estimated and, in many cases, are drastically different from the
Standard Model predictions. In particular, $\apks$ and $\apnn$
are both much smaller than one.

The mechanism that is responsible for the approximate nature of CP
is basically the following. The full high energy theory has exact
horizontal and CP symmetries. There are three relevant high energy
scales: $\Lambda_H$, where the horizontal symmetry is spontaneously
broken; $\Lambda_{CP}$, where CP is spontaneously broken; and
$\Lambda_F$, where the information about these spontaneous breakings
are communicated to the observable sector. There exists a hierarchy
between these scales: $\Lambda_{CP}\ll\Lambda_H\ll\Lambda_F$
(the hierarchy $\Lambda_{CP}\ll\Lambda_H$ is naturally produced
by the scalar potential), so that in the low energy effective
theory, the horizontal symmetry and the CP symmetry
appear explicitly broken by small parameters:
\eqn\small{\epsH\sim{\Lambda_H\over\Lambda_F},\ \ \
\epsCP\sim{\Lambda_{CP}\over\Lambda_F},\ \ \ \epsCP\ll\epsH .}
This mechanism, while predicting phenomenology of CP violation that
is very different from the Standard Model, also solves both the flavor
and CP problems of supersymmetry.

Our models use the Froggatt-Nielsen mechanism
\ref\FrNi{C.D. Froggatt and H.B. Nielsen, Nucl. Phys. B147 (1979) 277.}\
to achieve the small breaking parameters. We employ supersymmetric
abelian horizontal symmetries similarly to
\ref\LNS{M. Leurer, Y. Nir and N. Seiberg,
Nucl. Phys. B398 (1993) 319, hep-ph/9212278.}.
The supersymmetric flavor problems are solved by the alignment mechanism
\nref\NiSe{Y. Nir and N. Seiberg, Phys. Lett. B309 (1993) 337,
hep-ph/9304307.}%
\nref\LNSb{M. Leurer, Y. Nir and N. Seiberg,
Nucl. Phys. B420 (1994) 468, hep-ph/9310320.}%
\refs{\NiSe,\LNSb}.
As concerns CP violation in the supersymmetric framework, the idea
of approximate CP has been discussed in refs.
\nref\Pomb{A. Pomarol, Phys. Rev. D47 (1993) 273, hep-ph/9208205.}%
\nref\BaBb{K.S. Babu and S.M. Barr, Phys. Rev. Lett. 72 (1994) 2831,
 hep-ph/9309249.}%
\nref\AbFr{S.A. Abel and J.M. Frere, Phys. Rev. D55 (1997) 1623,
 hep-ph/9608251.}%
\refs{\Pomb-\AbFr}, and spontaneous CP violation has been discussed in
refs.
\nref\DHR{A. Dannenberg, L. Hall and L. Randall,
 Nucl. Phys. B271 (1986) 574.}%
\nref\BaMa{S.M. Barr and A. Masiero, Phys. Rev. D38 (1988) 366.}%
\nref\Maek{N. Maekawa, Phys. Lett. B282 (1992) 387.}%
\nref\Poma{A. Pomarol, Phys. Lett. B287 (1992) 331, hep-ph/9205247.}%
\nref\BaSe{S.M. Barr and G. Segre, Phys. Rev. D48 (1993) 302.}%
\nref\BaBa{K.S. Babu and S.M. Barr, Phys. Rev. D49 (1994) R2156,
 hep-ph/9308217.}%
\nref\MaRa{M. Masip and A. Rasin, Phys. Rev. D52 (1995) 3768,
 hep-ph/9506471.}%
\nref\MaRb{M. Masip and A. Rasin, Nucl. Phys. B460 (1996) 449,
 hep-ph/9508365.}%
\nref\NiRa{Y. Nir and R. Rattazzi,
Phys. Lett. B382 (1996) 363, hep-ph/9603233.}%
\refs{\DHR-\NiRa,\Pomb-\BaBb}.
Our work is closely related to two of these works. In ref. \BaBb, models
were constructed with spontaneous CP breaking and approximate CP in the
low energy theory. However, while the mechanism of communicating the
breaking in ref. \BaBb\ is aimed to solve the strong CP problem and leads
to a single low energy phase, our mechanism is aimed to solve the
supersymmetric flavor problems and leads to a large number of
low energy phases. Both the breaking mechanism and the communication
mechanism are the same as in ref. \NiRa.
The main new ingredient in our models is that, while the models of
ref. \NiRa\ have effectively CP breaking parameters of order one,
the models presented in this work give small CP breaking and,
therefore, a very different phenomenology of CP violation.
Moreover, as the supersymmetric CP problem is solved partially
by the approximate nature of CP, the required alignment is much
less precise than in existing models. This situation gives more freedom
in constructing the models and also allows for some different
phenomenological signatures in FCNC processes.
(We do not consider the strong CP problem in this work. Note, however,
that the alignment models may solve this problem too
\ref\Barr{S.M. Barr, Phys. Rev. D56 (1997) 5761, hep-ph/9705265.}.)

The structure of this paper is as follows. We first present two
explicit models of approximate CP, one where the breaking parameter
is intermediate, $\O(0.04)$ (section 2) and the other where it is
very small, $\O(0.001)$ (section 3). The implications of these
models for flavor changing neutral current processes are studied
in section 4 and for CP violation in section 5. Section 6 clarifies
an interesting point about holomorphic zeros, which bears consequences
for rare $K$ decays. Our conclusions are summarized in section 7.

\newsec{Model I}

Our first model employs a horizontal symmetry
\eqn\Hone{H=U(1)_1\times U(1)_2.}
The superfields of the supersymmetric standard model (SSM)
carry the following $H$-charges:
\eqn\HSMone{\matrix{
Q_1(2,1)&Q_2(3,-1)&Q_3(0,0)&&
\bar d_1(4,-1)&\bar d_2(-2,4)&\bar d_3(1,1)\cr
\bar u_1(5,-1)&\bar u_2(-2,4)&\bar u_3(0,0)&&
\phi_d(-1,0)&\phi_u(0,0)\cr}}
where $Q_i$ are the quark doublets, $\bar d_i$ and $\bar u_i$ are
the down and up quark singlets, and $\phi_i$ are the Higgs doublet
fields. In addition, we have three standard model singlet superfields:
\eqn\HSone{S_1(-1,0),\ \ S_2(0,-1),\ \ S_3(-3,-1).}

The horizontal symmetry is spontaneously broken when the three
$S_i$ fields assume VEVs. The breaking scale is somewhat lower
than a scale $M$ where the information about this breaking is
communicated to the SSM, presumably by heavy quarks in vector-like
representations of the Standard Model \FrNi. We will quantify all
the small parameters as powers of a small parameter $\lambda$ which
we take to be of ${\cal O}(0.2)$. Then, we take for the three VEVs
\eqn\epsone{\epsa\equiv{\vev{S_1}\over M}\sim\lambda,\ \ \
\epsb\equiv{\vev{S_2}\over M}\sim\lambda,\ \ \
\epsc\equiv{\vev{S_3}\over M}\sim\lambda^4.}
Note that due to the $U(1)_1\times U(1)_2$ symmetry, we can always
choose $\vev{S_1}$ and $\vev{S_2}$ to be real. However, $\vev{S_3}$
is, in general, complex with a phase of $\O(1)$. Then CP is spontaneously
broken by $\epsc$. The hierarchy $\epsc\ll\epsa,\epsb$ and $\arg(\epsc)
=\O(1)$ can be naturally induced, as explained below.

The electroweak symmetry is spontaneously broken by the VEVs of
$\phi_d$ and $\phi_u$, and we assume that
\eqn\tanbone{\tan\beta\equiv{\vev{\phi_u}\over\vev{\phi_d}}\sim{1\over
\lambda^2}.}

The model is defined by the horizontal symmetry, the assigned horizontal
charges and the hierarchy of VEVs. For most of our purposes, however,
we need not consider the full high energy theory. It is sufficient to
analyze the effective low energy theory, which is the SSM
supplemented with selection rules that follow from the
$H$-breaking pattern:
\item{(a)} Terms in the superpotential that carry charge $(m,n)$
under $H$ with $m,n\geq0$ are suppressed by $\O(\lambda^{m+n})$,
while those with $m<0$ and/or $n<0$ are forbidden (due to the
holomorphy of the superpotential \LNS).
\item{(b)} Terms in the K\"ahler potential that carry charge $(m,n)$
under $H$ are suppressed by $\O(\lambda^{|m|+|n|})$.

These selection rules allow us to estimate the various
entries in the quark mass matrices $M^q$ and the squark mass-squared
matrices $\tilde M^{q2}$. For each entry, we write the leading
contribution and the subleading contribution if it is complex
with respect to the leading one (namely, if it has a different
$\epsc$-dependence). We do not write the coefficients of $\O(1)$
which appear in each entry. For the quark mass matrices and for the
off-diagonal blocks in the squark mass-squared matrices, we write
the effective matrices after the rotations needed to bring the kinetic
terms into their canonical form have been taken into account \LNSb.
We get:
\eqn\dmassone{
M^d\sim\vev{\phi_d}\pmatrix{
\epsa^5&\epsa\epsb^5&\epsa^2\epsb^2+\epsc\epsa\epsb\cr
\epsa^6\epsb^2+\epsc\epsa^3\epsb^3&\epsb^3&\epsa^3+\epsc\epsb^3\cr
\epsa^3\epsb^3+\epsc\epsb^4&\epsa^3\epsb^4+\epsc^*\epsb^5&\epsb\cr},}
\eqn\umassone{
M^u\sim\vev{\phi_u}\pmatrix{
\epsa^7+\epsc(\epsa^4\epsb^3+\epsa^6\epsb)
&\epsb^5&\epsa^2\epsb+\epsc\epsa\cr
\epsa^8\epsb^2+\epsc\epsa^5\epsb^3&\epsa\epsb^3&
\epsa^3\epsb+\epsc\epsb^2\cr  \epsa^5\epsb+\epsc\epsa^2\epsb^2&
\epsa^2\epsb^4+\epsc^*\epsa\epsb^5&1\cr},}
\eqn\LLone{
\tilde M^{q2}_{LL}\sim\tilde m^2\pmatrix{
1&\epsa\epsb^2&\epsa^2\epsb+\epsc\epsa\cr
\epsa\epsb^2&1&\epsa^3\epsb+\epsc\epsb^2\cr
\epsa^2\epsb+\epsc^*\epsa&\epsa^3\epsb+\epsc^*\epsb^2&1\cr},}
\eqn\dRone{
\tilde M^{d2}_{RR}\sim\tilde m^2\pmatrix{
1&\epsa^6\epsb^5+\epsc^*\epsa^3\epsb^6&\epsa^3\epsb^2+\epsc^*\epsb^3\cr
\epsa^6\epsb^5+\epsc\epsa^3\epsb^6&1&\epsa^3\epsb^3+\epsc\epsb^4\cr
\epsa^3\epsb^2+\epsc\epsb^3&\epsa^3\epsb^3+\epsc^*\epsb^4&1\cr},}
\eqn\uRone{
\tilde M^{u2}_{RR}\sim\tilde m^2\pmatrix{
1&\epsa^7\epsb^5+\epsc^*\epsa^4\epsb^6&
\epsa^5\epsb+\epsc^*\epsa^2\epsb^2\cr
\epsa^7\epsb^5+\epsc\epsa^4\epsb^6&1&\epsa^2\epsb^4+\epsc\epsa\epsb^5\cr
\epsa^5\epsb+\epsc\epsa^2\epsb^2&
\epsa^2\epsb^4+\epsc^*\epsa\epsb^5&1\cr},}
\eqn\LRone{(\tilde M^{q2}_{LR})_{ij}\sim\tilde m(M^q)_{ij}.}
We can also estimate the size of the bilinear $\mu$ and $B$ terms:
\eqn\muone{\eqalign{\mu\sim&\tilde m(\epsa+\epsc^*\epsa^2\epsb),\cr
m_{12}^2\sim&\tilde m^2(\epsa+\epsc^*\epsa^2\epsb).\cr}}
Thus the horizontal symmetry solves the $\mu$-problem in the
way suggested in ref.
\ref\nirmu{Y. Nir, Phys. Lett. B354 (1995) 107, hep-ph/9504312.}.

From the mass matrices, we can further estimate the mixing angles
in the CKM matrix and in the gaugino couplings to quarks and squarks.
We denote the latter by $K^q_M$ where, for example, $K^d_L$ is
the mixing matrix that describes the gluino couplings to left-handed
down quarks and `left-handed' down squarks. (The LR mixing angles are
very small and we do not present them explicitly.)
We write the estimates in terms of powers of $\lam$.
For the CKM matrix, we find
\eqn\CKMone{|V_{us}|\sim\lam,\ \ \
|V_{ub}|\sim\lam^3,\ \ \ |V_{cb}|\sim\lam^2,}
as required by direct measurements, and
\eqn\dkmone{\dkm\sim\lam^2.}
For the gaugino couplings we find
\eqn\dLLone{(K^d_L)_{12}\sim\lam^3e^{i\lam^4},\ \ \
(K^d_L)_{13}\sim\lam^3e^{i\lam^2},\ \ \
(K^d_L)_{23}\sim\lam^2e^{i\lam^4},}
\eqn\uLLone{(K^u_L)_{12}\sim\lam,\ \ \
(K^u_L)_{13}\sim\lam^3e^{i\lam^2},\ \ \
(K^u_L)_{23}\sim\lam^4e^{i\lam^2},}
\eqn\dRRone{(K^d_R)_{12}\sim\lam^5e^{i\lam^2},\ \ \
(K^d_R)_{13}\sim\lam^5e^{i\lam^2},\ \ \
(K^d_R)_{23}\sim\lam^4e^{i\lam^4},}
\eqn\uRRone{(K^u_R)_{12}\sim\lam^4e^{i\lam^4},\ \ \
(K^u_R)_{13}\sim\lam^6e^{i\lam^2},\ \ \
(K^u_R)_{23}\sim\lam^6e^{i\lam^4}.}
Note that in \dLLone-\uRRone\ we omit coefficients of order one not
only in the overall magnitude but also in the phases.

Finally, we can estimate the relevant supersymmetric CP violating phases
\nref\DGH{M. Dugan, B. Grinstein and L. Hall,
 Nucl. Phys. B255 (1985) 413.}%
\nref\DiTh{S. Dimopoulos and S. Thomas, Nucl. Phys. B465 (1996) 23,
 hep-ph/9510220.}%
\refs{\DGH,\DiTh}:
\eqn\phiBone{\phi_B\equiv\arg(m_{12}^2/\mu)\sim\lam^6,}
\eqn\phiAone{\phi_A^u\equiv\arg\left({[V_L^u M^u V_R^{u\dagger}]_{11}
\over[V_L^u \tilde M^{u2}_{LR} V_R^{u\dagger}]_{11}}\right)\sim\lam^4,}
while the corresponding $\phi_A^d$ is negligible.

Before concluding this section, we would like to show how a complex
$\vev{S_3}$ which is hierarchically smaller than $\vev{S_1}$ and
$\vev{S_2}$ can be achieved naturally. Let us add yet another
Standard Model singlet field $S_4(6,2)$. The $S_i$ dependent terms
in the superpotential are
\eqn\WSone{W(S_i)\sim {a\over M^6}S_4S_1^6S_2^2
+{b\over M^3}S_4S_1^3S_2S_3+c S_4S_3^2,}
where $a,b,c$ are dimensionless numbers of $\O(1)$. For $\vev{S_4}=0$
we have $F_{S_1}=F_{S_2}=F_{S_3}=0$, while $F_{S_4}=0$ requires
\eqn\minone{a\epsa^6\epsb^2+b\epsa^3\epsb\epsc+c\epsc^2=0\ \
\Longrightarrow\ \ {\epsc\over\epsa^3\epsb}={-b\pm\sqrt{b^2-4ac}\over2c}.
}
We see that indeed $|\epsc|\sim|\epsa^3\epsb|\sim\lam^4$ and that
for $b^2-4ac<0$, $\epsc$ is complex.
(This mechanism for spontaneously breaking CP was first suggested
in ref. \BaBb.)

\newsec{Model II}

Our second model employs a horizontal symmetry
\eqn\Htwo{H=U(1)_1\times U(1)_2\times U(1)_3.}
The SSM superfields carry the following $H$-charges:
\eqn\HSMtwo{\matrix{
Q_1(2,0,1)&Q_2(0,0,2)&Q_3(0,0,0)&&
\bar d_1(0,5,0)&\bar d_2(1,5,-2)&\bar d_3(4,0,0)\cr
\bar u_1(-2,6,0)&\bar u_2(1,1,0)&\bar u_3(0,0,0)&&
\phi_d(-1,0,0)&\phi_u(0,0,0).\cr}}
We have four standard model gauge singlet fields:
\eqn\HStwo{S_1(-1,0,0),\ \ S_2(0,-1,0),\ \ S_3(0,0,-1),\ \ S_4(0,0,-4).}
The orders of magnitude of the various $S$-VEVs are
\eqn\epstwo{\epsa\equiv{\vev{S_1}\over M}\sim\lam,\ \ \
\epsb\equiv{\vev{S_2}\over M}\sim\lam,\ \ \
\epsc\equiv{\vev{S_3}\over M}\sim\lam,\ \ \
\epsd\equiv{\vev{S_4}\over M}\sim\lambda^4.}
Due to the $U(1)_1\times U(1)_2\times U(1)_3$ symmetry, we can always
choose $\vev{S_1}$, $\vev{S_2}$, $\vev{S_3}$ real, but $\vev{S_4}$
is, in general, complex with a phase of $\O(1)$. For the
electroweak breaking VEVs, we take
\eqn\tanbtwo{\tan\beta\sim1.}

For the various quark and squark mass matrices, we get:
\eqn\dmasstwo{
M^d\sim\vev{\phi_d}\pmatrix{
\epsa\epsb^5\epsc&\epsa^2\epsb^5\epsc(1+\epsc^2\epsd^*)
&\epsa^5\epsc(1+\epsc^2\epsd)\cr
\epsa\epsb^5\epsc^2(1+\epsd)&\epsb^5&\epsa^3\epsc^2(1+\epsd)\cr
M^d_{31}&\epsb^5\epsc^2(1+\epsd^*)&\epsa^3\cr},}
where $M^d_{31}\sim\epsa\epsb^5\epsc^2(\epsa^2+\epsc^2(1+\epsd
+\epsd^*))$,
\eqn\umasstwo{
M^u\sim\vev{\phi_u}\pmatrix{
\epsb^6\epsc&\epsa^3\epsb\epsc(1+\epsc^2\epsd)&
\epsa^2\epsc(1+\epsc^2\epsd)\cr
\epsa^2\epsb^6\epsc^2(1+\epsd)&\epsa\epsb\epsc^2(1+\epsd)&
\epsc^2(1+\epsd)\cr \epsa^2\epsb^6&\epsa\epsb&1\cr},}
\eqn\LLtwo{
\tilde M^{q2}_{LL}\sim\tilde m^2\pmatrix{
1&\epsa^2\epsc(1+\epsc^2\epsd^*)&\epsa^2\epsc(1+\epsc^2\epsd)\cr
\epsa^2\epsc(1+\epsc^2\epsd)&1&\epsc^2(1+\epsd)\cr
\epsa^2\epsc(1+\epsc^2\epsd^*)&\epsc^2(1+\epsd^*)&1\cr},}
\eqn\dRtwo{
\tilde M^{d2}_{RR}\sim\tilde m^2\pmatrix{
1&\epsa\epsc^2(1+\epsd^*)&\epsa^4\epsb^5\cr
\epsa\epsc^2(1+\epsd)&1&\epsa^3\epsb^5\epsc^2(1+\epsd)\cr
\epsa^4\epsb^5&\epsa^3\epsb^5\epsc^2(1+\epsd^*)&1\cr},}
\eqn\uRtwo{
\tilde M^{u2}_{RR}\sim\tilde m^2\pmatrix{
1&\epsa^3\epsb^5&\epsa^2\epsb^6\cr
\epsa^3\epsb^5&1&\epsa\epsb\cr
\epsa^2\epsb^6&\epsa\epsb&1\cr},}
\eqn\LRtwo{(\tilde M^{q2}_{LR})_{ij}\sim\tilde m(M^q)_{ij}.}
For the bilinear terms, we find
\eqn\mutwo{\eqalign{\mu\sim&\tilde m\epsa(1+\epsc^4(\epsd+\epsd^*)),\cr
m_{12}^2\sim&\tilde m^2\epsa(1+\epsc^4(\epsd+\epsd^*)).\cr}}

For the CKM matrix, we find again magnitudes consistent with
the measurements (namely, the same orders of magnitude as in \CKMone)
but the KM phase is smaller:
\eqn\dkmtwe{\dkm\sim\lam^4.}
For the gaugino couplings we find
\eqn\dLLtwo{(K^d_L)_{12}\sim\lam^3e^{i\lam^6},\ \ \
(K^d_L)_{13}\sim\lam^3e^{i\lam^6},\ \ \
(K^d_L)_{23}\sim\lam^2e^{i\lam^4},}
\eqn\uLLtwo{(K^u_L)_{12}\sim\lam e^{i\lam^4},\ \ \
(K^u_L)_{13}\sim\lam^3e^{i\lam^4},\ \ \
(K^u_L)_{23}\sim\lam^2e^{i\lam^4},}
\eqn\dRRtwo{(K^d_R)_{12}\sim\lam^3e^{i\lam^4},\ \ \
(K^d_R)_{13}\sim\lam^7e^{i\lam^4},\ \ \
(K^d_R)_{23}\sim\lam^4e^{i\lam^4},}
\eqn\uRRtwo{(K^u_R)_{12}\sim\lam^4e^{i\lam^6},\ \ \
(K^u_R)_{13}\sim\lam^6e^{i\lam^6},\ \ \
(K^u_R)_{23}\sim\lam^2.}

The supersymmetric CP violating phases are:
\eqn\phiBtwo{\phi_B\sim\lam^8,}
\eqn\phiAtwo{\phi_A^u\sim\lam^6,}
while $\phi_A^d$ is negligible.

The required hierarchy between the $H$ and CP breaking scales is
achieved by minimizing the Higgs potential for the four $S_i$ of eq.
\HStwo\ and a fifth singlet field $S_5(0,0,8)$. This would give
$\epsd=\O(\epsc^4)$ and complex.

\newsec{Flavor Changing Neutral Current Processes}

Generic supersymmetric models, with mass-squared differences between
generations of $\O(\tilde m^2)$ ($\tilde m$ is the supersymmetry
breaking scale) and supersymmetric mixing angles of $\O(1)$ give much too
large contributions to various flavor changing neutral current (FCNC)
processes such as $\Delta m_K$, $\Delta m_D$ and $\Delta m_B$.
There are various solutions to this problem:
\item{a.} All squark generations are equal at some high energy scale.
This is the situation, for example, in models of gauge mediated
supersymmetry breaking.
\item{b.} The first two squark generations are degenerate due to
a non-Abelian horizontal symmetry.
\item{c.} The first two squark generations are very heavy.
\item{d.} Squarks are neither degenerate nor very heavy, but the mixing
angles in the gaugino couplings to quarks and squarks are small.

The last option arises naturally in models of Abelian horizontal
symmetries of the type that we used in constructing our models.
Indeed, one can easily see from eqs. \dLLone--\uRRone\ and
\dLLtwo--\uRRtwo\ that there is no mixing angle of order one in our
models; they are all suppressed by the selection rules of the
horizontal symmetries. To understand whether the alignment in
our models is precise enough to satisfy the phenomenological constraints
and, in the case that it is, whether the supersymmetric contributions
are significant in comparison to the Standard Model ones, we
write down the constraints on the mixing angles (taken from ref.
\ref\GGMS{F. Gabbiani, E. Gabrielli, A. Masiero and L. Silvestrini,
Nucl. Phys. B477 (1996) 321, hep-ph/9604387.})
in terms of powers of $\lam$ and then compare to the predictions
of our two models. This is done in Table 1. (We define
$\vev{K_{ij}}\equiv[(K_L)_{ij}(K_R)_{ij}]^{1/2}$.)
\vskip 0.7cm

\begintable
Mixing Angle & Process & Bound & Model I & Model II \crthick
$(K^d_L)_{12}$ & $\Delta m_K$ & $\lam-\lam^2$ & $\lam^3$ & $\lam^3$ \nr
$(K^d_R)_{12}$ & $\Delta m_K$ & $\lam-\lam^2$ & $\lam^5$ & $\lam^3$ \nr
$\vev{K^d_{12}}$ & $\Delta m_K$ & $\lam^3$ & $\lam^4$ & $\lam^3$ \nr
$(K^d_L)_{13}$ & $\Delta m_B$ & $\lam$ & $\lam^3$ & $\lam^3$ \nr
$(K^d_R)_{13}$ & $\Delta m_B$ & $\lam$ & $\lam^5$ & $\lam^7$ \nr
$\vev{K^d_{13}}$ & $\Delta m_B$ & $\lam^2$ & $\lam^4$ & $\lam^5$ \nr
$(K^u_L)_{12}$ & $\Delta m_D$ & $\lam$ & $\lam$ & $\lam$ \nr
$(K^u_R)_{12}$ & $\Delta m_D$ & $\lam$ & $\lam^4$ & $\lam^4$ \nr
$\vev{K^u_{12}}$ & $\Delta m_D$ & $\lam^2$ & $\lam^{5/2}$ & $\lam^{5/2}$
\endtable

\centerline{Table 1. Supersymmetric mixing angles in our models and the
phenomenological bounds on them.}

We learn the following points from the Table:

(i) The contributions to $\Delta m_D$ that are proportional to
$[(K^u_L)_{12}]^2$ saturate the experimental upper bound in both
models. This is a generic feature of models of alignment
\refs{\NiSe,\LNSb}, related to the fact that in these models
the Cabibbo mixing ($|V_{us}|\sim\lam$) comes from the up sector.

(ii) The contributions to $\Delta m_B$ are very small. In all
alignment models, the standard model amplitudes dominate \LNSb.
But while the supersymmetric contributions could be generically of
$\O(20\%)$, the models constructed here provide an example where
these contributions are below the percent level.

(iii) The contributions to $\Delta m_K$ are of $\O(10\%)$ in model I
and saturate the experimental value for model II. This is in contrast
to all previous models of alignment where, to satisfy the $\epsK$
constraint, the supersymmetric contributions to $\Delta m_K$ were
negligibly small. The large contribution comes in the two models from
$(K_L^d)_{12}(K_R^d)_{12}$.

Before proceeding, we would like to make two comments:
\item{a.} The contributions to other FCNC processes, such as
$\Delta m_{B_s}$ and $b\ra s\gamma$, are very small. The
$K^+\ra\pi^+\nu\bar\nu$ decay is discussed separately below.
\item{b.} The contributions from the (LR) blocks in the squark
mass-squared matrices are much smaller than those coming from the
mixing angles presented in Table 1. This is the reason why, even though
we calculated them explicitly, we do not present them in Table 1.

As concerns the rare $K^+\ra\pi^+\nu\bar\nu$ decay, the largest
supersymmetric contribution in alignment models comes from
$(K^d_L)_{12}$, if it is as large as allowed by the $\Delta m_K$
constraint, $(K^d_L)_{12}\sim\lam^2$
\ref\NiWo{Y. Nir and M.P. Worah, hep-ph/9711215.}.
In such a case, the supersymmetric contributions from penguin diagrams
with chargino and $\tilde u,\tilde c$ squarks are significant.
In both our models, $(K^d_L)_{12}\sim\lam^3$,
leading to supersymmetric contributions of $\O(10\%)$. While both the
standard model and the supersymmetric amplitudes are real to a good
approximation, so that there is maximal interference between the two,
the relative sign is unknown so that the rate could be either enhanced
or suppressed compared to the standard model.

It is interesting that we are unable to construct a model where either
$(K_L^d)_{12}$ or $(K_R^d)_{12}$ are as large as allowed, namely
$\O(\lam^2)$. This situation goes beyond the two specific models
that we present here and seems generic to models with continuous
Abelian horizontal symmetries. The reason for that is explained in
section 7. The situation is different in models of {\it discrete}
Abelian symmetries. We actually constructed a model with a
$Z_4\times Z_9$ horizontal symmetry where $(K_L^d)_{12}\sim\lam^2$.
The model is, however, quite complicated and its Higgs potential
does not provide in a natural way the hierarchy between $\Lambda_H$
and $\Lambda_{CP}$ (this seems a rather generic feature of models
with discrete Abelian symmetries), which is the reason that we do
not present it here.

The contribution to $K^+\ra\pi^+\nu\bar\nu$ from the (LR) sector
is negligibly small. For the (LR) contributions to be significant,
we need the off-diagonal terms in $\tilde M^{q2}_{LR}$ to be of
$\O(m_t\tilde m)$
\ref\BRS{A.J. Buras, A. Romanino and L. Silvestrini, hep-ph/9712398.},
while in our models they are much smaller than that, as can be seen from
eqs. \LRone\ and \LRtwo.

\newsec{CP Violation}

Each of the two models that we have constructed has an approximate
CP symmetry for the SSM. In model I, all CP violating phases are
$\leq\O(\lam^2)$ and in model II, they are $\leq\O(\lam^4)$.
The resulting predictions for CP violating observables are then
drastically different from the Standard Model.

The first thing to note is that with $\dkm\sim\lam^2$ or $\lam^4$,
it is impossible to account for $\epsK\sim10^{-3}$ by the
Standard Model contributions. However, in both models,
\eqn\Imonetwo{\Im[(K^d_L)_{12}(K^d_R)_{12}]\sim\lam^{10},}
which can account for $\epsK$ from the supersymmetric gluino-mediated
diagrams.

The most dramatic consequences of the approximate CP symmetry
concern the CP violating asymmetries that are expected to be
large in the standard model. First, let us consider CP asymmetries
in neutral $B$ decays into final CP eigenstates. For the sake of
definiteness, we consider $\apks$. The supersymmetric contributions
to the $B-\bar B$ mixing amplitude are, as mentioned above,
negligible. Usually this leads to the conclusion that the
standard model predictions for $\apks$ remain valid. But this is
definitely not the case in our framework. The fact that $\dkm$
is very small means that so will be $\apks$. Explicitly,
\eqn\apksaCP{{\apks\over\apks^{\rm SM}}=\cases{
\O(\lam^2)&Model I,\cr \O(\lam^4)&Model II.\cr}}
If $\apks$ is measured to be in the Standard Model
range, our models of approximate CP will be excluded.

Concerning $\apnn$, it was shown that $\apnn=\sin^2\theta_K$, where
$\theta_K$ is the relative phase between the $K-\bar K$ mixing
amplitude and the $s\ra d\nu\bar\nu$ decay amplitude
\ref\GrNi{Y. Grossman and Y. Nir, Phys. Lett. B398 (1997) 163.}.
In our models, the standard model contributions dominate
both the real part and the imaginary part of the decay amplitude.
In model I, the standard model also dominates the real part of the
mixing amplitude, while the supersymmetric contribution dominates
the imaginary part. In model II, the two contributions to the real
part are comparable, but supersymmetric diagrams dominate the imaginary
part of the mixing amplitude. In either case, the approximate CP
symmetry leads to a strong suppression of $\apnn$:
\eqn\apnnaCP{{\apnn\over\apnn^{\rm SM}}=\cases{
\O(\lam^4)&Model I,\cr \O(\lam^8)&Model II.\cr}}
We learn that if $\apnn$ is measured in the foreseeable future,
our models of approximate CP will be excluded.

CP could play an interesting role in $D-\bar D$ mixing
\nref\BSN{G. Blaylock, A. Seiden and Y. Nir,
 Phys. Lett. B355 (1995) 555, hep-ph/9504306.}%
\nref\Wolf{L. Wolfenstein, Phys. Rev. Lett. 75 (1995) 2460,
 hep-ph/9505285.}%
\refs{\BSN,\Wolf}. In measuring the time-dependent decay rate for
$D^0\ra K^+\pi^-$ and $\bar D^0\ra K^-\pi^+$, a term proportional to
$te^{-\Gamma t}$ that is different between the two CP-conjugate modes
will appear if the relative phase between the $D-\bar D$ mixing
amplitude and the $c\ra u\bar s d$ decay amplitude, $\theta_D=\arg
\lam_{D^0\ra K^+\pi^-}$, is large.
This is the case in previous alignment models. However,
in our models, the relevant phase, that is $\arg[(K^u_L)_{12}^2]$,
is very small ($\leq\O(\lam^4)$), so the effect is probably unobservable.
If $D-\bar D$ mixing is not observed within, say,
one order of magnitude of the present experimental bound, then
the existing alignment models are excluded. But if such mixing is
observed and with large CP violation, then the alignment mechanism
remains viable but not in combination with approximate CP.

Finally, we discuss the electric dipole moment of the neutron $d_N$.
It was argued in ref. \GNR\
that in supersymmetric models without universality, namely when
there is no super-GIM mechanism, there is a generic lower bound on the
CP violating phases that contribute to $d_N$. This bound is of
$\O(\lam^6)$ and leads to $d_N\gsim10^{-28}\ e\ {\rm cm}$. This bound
is about three orders of magnitude above the value in supersymmetric
models with universality and may be within the reach of forthcoming
experiments. Indeed, our models obey this bound and predict a
potentially observable $d_N$.

Our results concerning CP violation are summarized in Table 2.
Note that, within the Standard Model, $\apnn=\O(\sin^2\beta)$,
which parametrically is of $\O(1)$, but turns out to be numerically
of $\O(\lam)$
\ref\BBkp{G. Buchalla and A. Buras, Phys. Rev. D54 (1996) 6782,
 hep-ph/9607447.}.
$d_n$ is given in units of $10^{-23}\ e$ cm, so that the present
experimental bound is $d_N\lsim\lam^2$.
\vskip 1cm

\begintable
Process & SM & Model I & Model II \cr
$\apks$ & $\O(1)$ & $\O(\lam^2)$ & $\O(\lam^4)$ \nr
$\apnn$ & $1$ & $\O(\lam^4)$ & $\O(\lam^8)$ \nr
$\theta_D$ & 0 & $\ll\O(\lam^4)$ & $\O(\lam^4)$ \nr
$d_N$ & 0 & $\O(\lam^4)$ & $\O(\lam^6)$ \endtable

\centerline{Table 2. CP violating observables in the SM and in our
models.}

\newsec{Lifting Holomorphic Zeros}

The Yukawa couplings, being part of the superpotential, are
holomorphic in the $H$-breaking parameters. In particular,
if all breaking parameters carry charges of the same sign
under one of the horizontal $U(1)$'s, then an entry in the
Yukawa matrix that breaks $H$ by a charge of the same sign
vanishes \NiSe. We call these vanishing Yukawa couplings
`holomorphic zeros'. However, in the basis where these holomorphic
zeros are exact, the kinetic terms are not canonical. When we normalize
them back to a canonical form, the holomorphic zeros are lifted
\nref\DPS{E. Dudas, S. Pokorski and C.A. Savoy,
 Phys. Lett. B356 (1995) 45, hep-ph/9504292.}%
\refs{\LNSb,\DPS}.

Knowing the CKM mixing angles and the quark mass ratios allows us to
guess a `naive value' for each entry in the Yukawa matrices \LNS.
These are
\eqn\naive{Y^d\sim\lam^3\tan\beta\pmatrix{\lam^4&\lam^3&\lam^3\cr
\lam^3&\lam^2&\lam^2\cr \lam&1&1\cr},\ \ \ Y^u\sim\pmatrix{
\lam^7&\lam^5&\lam^3\cr \lam^6&\lam^4&\lam^2\cr \lam^4&\lam^2&1\cr},}
(where we have taken the high energy values $m_b/m_t\sim\lam^3$ and
$m_c/m_t\sim\lam^4$). When one of these entries vanishes because
of holomorphy, the rotation to bring the kinetic terms back to
the canonical normalization cannot lift this zero to its
naive value \LNSb. In fact, in our model, the entry in the effective
Yukawa matrix (that is, in the basis where the kinetic terms are
canonically normalized) is suppressed by at least $\lam^2$ compared
to the naive value. Below we prove this statement.

Before we provide the proof, we would like to point out a
phenomenological consequence of this situation. The naive value
$Y^d_{12}/Y^d_{22}\sim\sin\theta_C\sim\lam$ is
too large for the $\Delta m_K$ constraint (if there is no squark
degeneracy). The way that alignment models solve this problem is
by assuming that $Y^d_{12}$ is a holomorphic zero.
The $\Delta m_K$ constraint allows $Y^d_{12}/Y^d_{22}\sim\lam^2$.
If this bound were saturated, then the supersymmetric contribution
to the $K^+\ra\pi^+\nu\bar\nu$ decay could be significant
\NiWo. What we learn, however, is that the
maximal value that we can obtain for this ratio in our framework is
$Y^d_{12}/Y^d_{22}\sim\lam^3$ (which is, indeed, realized in both
models presented above). Therefore, in alignment models that employ
continuous Abelian horizontal symmetries with small breaking parameters
that are $\leq\O(\lam)$, the modification to the Standard Model
prediction for $K^+\ra\pi^+\nu\bar\nu$ is never large.

To prove that a lifted holomorphic zero is suppressed by at least
the square of the breaking parameters, let us consider a horizontal
symmetry $H=U(1)_x\times U(1)_y$ broken by small parameters of order
\eqn\breakthr{\epsilon_x(-1,0)\sim\lam^{n_x},\ \ \
\epsilon_y(0,-1)\sim\lam^{n_y},}
with two down-quark generations,
\eqn\toythr{Q_1(a_x,a_y),\ \ \bar d_1(b_x,b_y),\ \ Q_2(c_x,c_y),\ \
\bar d_2(d_x,d_y).}
The kinetic terms for the ($Q_1,Q_2$) fields
have coefficients of order
\eqn\KofQ{\pmatrix{1&\lam^{n_x|a_x-c_x|+n_y|a_y-c_y|}\cr
\lam^{n_x|a_x-c_x|+n_y|a_y-c_y|}&1\cr},}
while those of the ($\bar d_1,\bar d_2$) fields will be
\eqn\Kofd{\pmatrix{1&\lam^{n_x|b_x-d_x|+n_y|b_y-d_y|}\cr
\lam^{n_x|b_x-d_x|+n_y|b_y-d_y|}&1\cr}.}
We further assume, without loss of generality, that the Yukawa
coupling $Y^d_{12}$, which breaks $H$ by charge $(a_x+d_x,a_y+d_y)$,
is a holomorphic zero because $a_x+d_x<0$,
while the other entries in $Y^d$ do not vanish:
\eqn\Ydthr{Y^d\sim\pmatrix{\lam^{n_x(a_x+b_x)+n_y(a_y+b_y)}&0\cr
\lam^{n_x(c_x+b_x)+n_y(c_y+b_y)}&
\lam^{n_x(c_x+d_x)+n_y(c_y+d_y)}\cr}.}

A straightforward calculation shows then that the effective $Y^d_{12}$ is
\eqn\effYd{\eqalign{(Y^d_{12})_{\rm eff}=&\ A+B+C,\cr
A=&\ \lam^{n_x(a_x+b_x)+n_y(a_y+b_y)+n_x|b_x-d_x|+n_y|b_y-d_y|}\cr
B=&\ \lam^{n_x(c_x+d_x)+n_y(c_y+d_y)+n_x|a_x-c_x|+n_y|a_y-c_y|}\cr
C=&\ \lam^{n_x(b_x+c_x)+n_y(b_y+c_y)+n_x(|a_x-c_x|+|b_x-d_x|)
+n_y(|a_y-c_y|+|b_y-d_y|)}.\cr}}
This should be compared to the `naive' value, which is
\eqn\naiYd{(Y^d_{12})_{\rm naive}\sim\lam^{n_x(a_x+d_x)+n_y(a_y+d_y)}.}
Let us first examine $A$ of eq. \effYd.
We assume that $d_y\geq b_y$ (otherwise the suppression is even
stronger). From $a_x+b_x\geq0$ and $a_x+d_x<0$, we conclude that
$b_x>d_x$. We learn that
\eqn\Anaive{{A\over(Y^d_{12})_{\rm naive}}\sim
\lam^{2n_x(b_x-d_x)}\leq\lam^{2n_x}.}
Second, we examine $B$ of eq. \effYd. We assume that $a_y\geq c_y$.
From $c_x+d_x\geq0$, we conclude that $c_x>a_x$. We learn that
\eqn\Bnaive{{B\over(Y^d_{12})_{\rm naive}}\sim
\lam^{2n_x(c_x-a_x)}\leq\lam^{2n_x}.}
As concerns $C$, it is easy to see that it is smaller than $A$ and $B$.

The final conclusion is then as follows. Suppose that a holomorphic
zero is induced because the Yukawa coupling carries a negative
charge under a symmetry $U(1)_x$ that is broken by a small parameter
$\epsilon_x\sim\lam^{n_x}$. Then, the effective Yukawa coupling, that
is the coupling in the basis where the kinetic terms are canonically
normalized, obeys
\eqn\effnai{{(Y^q_{ij})_{\rm eff}\over
(Y^q_{ij})_{\rm naive}}\lsim\lam^{2n_x}.}
Since in all our models $n_x\geq1$, the suppression is at least
by $\O(\lam^2)$.

\newsec{Conclusions}

Supersymmetry allows for CP violating mechanisms that are dramatically
different from the Standard Model. In particular, CP could be an
approximate symmetry, with all CP violating phases very small,
$$10^{-3}\lsim\phi_{CP}\ll1.$$
In this work, we gave two examples of phenomenologically viable models
where CP is broken by paramaters of order $0.04$ or $0.001$.

The two models that we presented here are not unique. We use them
to demonstrate how approximate CP can arise naturally and
to explore the phenomenological signatures of approximate CP.
The specific models should only be thought of as
examples, but the underlying CP breaking mechanism and the
resulting phenomenological implications are generic to this class
of models.

The fact that the Standard Model and the models of approximate
CP are both viable at present is related to the fact that the
mechanism of CP violation has not really been tested experimentally.
The only measured CP violating observale, that is $\epsK$, is small.
Its smallness could be related to the
`accidental' smallness of CP violation
for the first two quark generations, as is the case in the Standard
Model, or to CP being an approximate symmetry, as is the case in the
models discussed here. Future measurements, particularly of processes
where the third generation plays a dominant role (such as $\apks$
or $\apnn$), will easily distinguish between the two scenarios.
While the Standard Model predicts large CP violating effects
for these processes, approximate CP would suppress them too.

The distinction between the Standard Model and Supersymmetry could
also be made -- though less easily -- in measurements of CP violation
in neutral $D$ decays and of the electric dipole moments of the
neutron. Here, the GIM mechanism of the Standard Model is so
efficient that CP violating effects are unobservable in both cases.
In contrast, the flavor breaking in supersymmetry might be much
stronger, and then the approximate CP somewhat suppresses the effects
but to a level which is perhaps still observable.

Finally, we note that the predictions for FCNC processes are modified
even for those processes where the supersymetric contribution is
negligible. The reason is that the constraints on the CKM parameters
are modified. Instead of the $\epsK$ constraint, which is not relevant
for the CKM parameters, we have $\eta\approx0$. The resulting
modifications were recently analyzed in ref.
\ref\BHSW{R. Barbieri, L. Hall, A. Stocchi and N. Weiner,
 hep-ph/9712252.}.

\vskip 1 cm
\centerline{\bf Acknowledgements}
Y.N. is supported in part by the United States -- Israel Binational
Science Foundation (BSF), by the Israel Science Foundation,
and by the Minerva Foundation (Munich).

\listrefs
\end